\documentclass[preprint,12pt]{elsarticle}
\usepackage{amssymb}

\journal{Journal of Information Security and Applications}

\begin{document}

\begin{frontmatter}

\title{Combining ensemble methods and social network metrics for improving the accuracy of OCSVM on intrusion detection in SCADA systems}

\author[rvt3]{Leandros A.  Maglaras\corref{cor2}}
\ead{l.maglaras@surrey.ac.uk}
\author[rvt]{Jianmin Jiang}
\ead{jianmin.jiang@surrey.ac.uk}
\author[rvt2]{Tiago J. Cruz}
\ead{tjcruz@dei.uc.pt}

\cortext[cor2]{Principal corresponding author}
\address[rvt3]{School of Computer Science and Informatics, De Montfort University, Leicester, U.K.}
\address[rvt]{Department of Computing, University of Surrey, Guildford, Surrey, U.K.}
\address[rvt2]{Department of Informatics Engineering, University of Coimbra, 15780, Portugal}

\title{Combining ensemble methods and social network metrics for improving accuracy of OCSVM on intrusion detection in SCADA systems}

\begin{abstract}
Modern Supervisory Control and Data Acquisition SCADA systems used by the electric utility industry to monitor and 
control electric power generation, transmission and distribution are recognized today as critical components of the electric power delivery infrastructure. 
SCADA systems are large, complex and incorporate increasing numbers of widely distributed components.
The presence of a real time intrusion detection mechanism, which can cope with different types of attacks, is of great importance, in order to defend a system against cyber attacks 
This defense mechanism must be distributed, cheap and above all accurate, since false positive alarms, or mistakes regarding the origin of the intrusion mean severe costs for the system. 
Recently an integrated detection mechanism, namely {\it IT-OCSVM} was proposed, which is distributed in a SCADA network as a part of a distributed intrusion detection system (IDS), providing accurate data about the origin and the time of an intrusion.
In this paper we also analyze the architecture of the integrated detection mechanism and we perform extensive simulations based on real cyber attacks in a small SCADA testbed in order to evaluate the performance of the proposed mechanism.

\end{abstract}


\end{frontmatter}

\section{Introduction}\label{sec-intro}

Cyber-physical systems are becoming vital for modernizing national critical infrastructure systems. Cyber attacks often target valuable infrastructure assets, taking advantage of architectural/technical vulnerabilities or even weaknesses in defense systems. Most weaknesses in CIs arise from the fact that many adopt off-the-shelf technologies from the IT world, without a significant change in terms of the operator mindset, thus remaining based on the "airgap" security principle that falsely assumes that an apparently isolated and obscure system is implicitly secure. Consequently, once a system is open to receiving off-the-shelf solutions, this increases its exposure to cyber-attacks. 

The proliferation of new technologies, especially Internet-like communications networks, may introduce some new threats to the security of a smart grid. In such a network there are three crucial aspects of security that may be threatened due to the 
CIA-triad, these being: confidentiality, integrity, and availability \cite{4781067}. 
Confidentiality  is  the property  that  information  is not made  available or disclosed  to unauthorized  individuals,  entities  or  processes.  An attack on this occurs when  an unauthorized person,  entity or process enters  the  system and  accesses the information.   
Integrity refers to safeguarding the accuracy and completeness of assets, which ensures that  the  information  in  the  system  will not  be modified  by attacks.
Availability pertains to the property of being accessible and usable upon demand by an authorized entity. The resources need to be kept accessible at all times to authorized entities or processes.

Beyond cyber threats like malware, spyware, and computer viruses that currently threaten the security of computer communication networks, the introduction of new and distributed technologies,  such  as  smart  meters,  sensors,  and  other  sub-networks  can  bring  new vulnerabilities to a smart grid \cite{5054916}.  In  the  three main  control  systems  of  a CI,  the  SCADA  the  central  nerve  system that  constantly gathers  the  latest  status  from  remote units. 
The  communication  system  for wide-area protection  and  control  of  a power  grid  can  be 
blocked or  cut off due  to  component  failures or  communication delays. If one of  the 
crucial  communication  channels  fails to  connect  in  the  operational  environment, 
inability to control or operate important facilities may occur with the possibility of power outages. 
In this situation, the effect of some widely known attacks can have devastating consequences on SCADA systems. 
Moreover, the design of SCADA systems is different from conventional IT networks, even when based on the same physical technology, such as Ethernet networks. That is, industrial control-specific protocols are used in SCADA systems, where a limited number packet types are exchanged between entities of the network.

Intrusion detection systems can be classified into centralized intrusion detection systems (CIDS) and distributed intrusion detection systems (DIDS) by the way in which their components are distributed \cite{balasubramaniyan1998architecture}. A CIDS is such that the analysis of the data will be performed  in  some  fixed  locations  without  considering  the  number  of  hosts  being monitored \cite{kumar1995classification}, while a DIDS is composed of several IDS over large networks whose data analysis  is performed  in  a  number  of  locations proportional  to  the  number  of  hosts. As one part of an intrusion detection system, the DIDS has specific advantages over the CIDS.  For instance, it  is highly  scalable  and  can provide  gradual degradation  of  service, easy extensibility and scalability \cite{crosbie1995active}. It is evident that the development of distributed IDS specifically designed for SCADA systems, being able to ensure an adequate balance between high accuracy, low false alarm rate and reduced network traffic overhead, is needed. The above discussion clearly indicates that specific intrusion detection systems that reassure both high accuracy, low rate of false alarms and decreased overhead on the network traffic need to be designed for SCADA systems.

\subsection{Motivation}\label{sec-moti}

Among other approaches, neural networks, support vector machines, K-nearest neighbor (KNN) and the hidden Markov model can be used for intrusion detection. While existing signature-based network IDS, such as Snort or Suricata can be effective in SCADA environments, they require specific customization for such a purpose. Also, they are not effective against rogue threats for which known patterns or signatures are not known. OCSVM principles have shown great potential in the area of anomaly detection \cite{wang2004anomaly,Sai2014}, being a natural extension of the support vector algorithm in the case of unlabeled data, especially for the detection of outliers.

Social network analysis ({\it SNA}) can be used in order to discover security policy breaches in a network and refers to the use of network theory to analyze social networks. That is, it views social relationships in terms of network theory, consisting of nodes, representing individual actors within the network, and ties which represent relationships between the individuals, such as friendships, kinships, organizations and sexual relationships. By using comparative metrics of network's structure during normal and abnormal operation,  we can discover security policy breaches in a network. One can assume that network communication between nodes, constitutes a social network of users and their applications, so the appropriate methods of social network formal analysis can be applied \cite{kolaczeksocial}. In on-line social systems perpetrators of malicious behavior often display patterns of interaction that are quite different from regular users, which can be identified through the application of anomaly detection techniques. Thus, in accordance  \cite{Gogoi01042011,aTeodoro200918}, network anomalies can be defined as patterns of interaction that significantly differ from the norm and in order to capture the appropriate patterns of interaction, specific aspects of entities' behavior are used (e.g. email analysis). Similar to this, in a SCADA system, individual entities demonstrate a quite different communication behavior when infected, in terms of mean packet generation frequency (traffic burst), protocol distribution or interaction pattern. 

Discovering anomalies in the context of a network system is a challenging issue due to the complexity of the environment and the different nature of the induced attacks. Regarding node behavior related decisions it makes sense to ask more than one decision mechanism, since this practice assures a more trusted final decision. Ensemble systems of classifiers are widely used for intrusion detection in networks \cite{curiac2012ensemble,shieh2009ensembles}. These aim to include mutually complementary individual classifiers, which are characterized by high diversity in terms of classifier structure \cite{tsoumakas2004effective}, internal parameters \cite{kim2010ensemble} or classifier inputs\cite{krawczyk2014diversity}.

In real time systems, in addition to fast response and accuracy limited communication between detection modules is also desirable. By sending an explicit message for every anomaly detected, the intrusion detection mechanism will flood the medium with messages that will cause a delay in the communication between entities in the SCADA system. Moreover, since the detection mechanism needs to be sited at several locations in the SCADA system in order to recognize the intrusion near the origin, the communication overhead caused by the detection mechanism is further increased.  One solution is the addition of a control channel, whereby these messages can be communicated without affecting the system's performance, but this is not always feasible. For this reason, an aggregation mechanism that groups initial alerts and sends a limited number of messages reporting the fault/intrusion accurately and on time is needed.

\subsection{Contributions}\label{sec-contr}
\label{subsec-contributions}

The present article analyzes and evaluates the performance of a recently proposed intrusion detection mechanism, namely the {\it IT-OCSVM}~\cite{Maglaras_IET_2014}, against the baseline OCSVM method.
The mechanism uses a central OCSVM and a cluster of automatically  produced ones, one for each source that induces significant traffic in the system, an embedded ensemble mechanism, social metrics, an aggregation method and a k-means clustering procedure that categorizes aggregated alerts. The mechanism runs in a distributed way and produces dedicated IDMEF (Intrusion Detection Message Exchange Format) messages that report the severity of the attack detected. The proposed mechanism is incorporated in a distributed   IDS (intrusion detection system) communicating with other detection and management components of the system. {\it IT-OCSVM} is evaluated using datasets extracted from a testbed that mimics a small scale SCADA system under normal and malicious operation. The evaluation of the proposed method attests the superiority of the new structure in terms of accuracy, false alarm rate and system overhead.

The rest of this article is organized as follows:
Section~\ref{sec-scada} presents the use of OCSVM in SCADA systems; 
Section~\ref{sec-itocsvm} describes the IT-OCSVM method;
Section~\ref{sec-evaluation} presents the simulation environment and results; an Section~\ref{sec-conclusions} 
concludes the article.

\section{OCSVM for intrusion detection in SCADA systems} \label{sec-scada}

\begin{figure}[!hbt]
\begin{center}
\includegraphics[width=0.65\textwidth,natwidth=420,natheight=400]{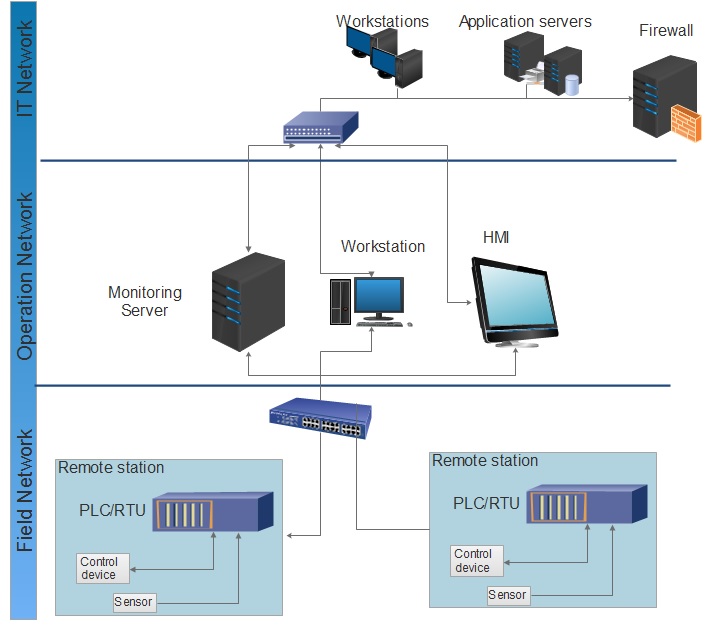}
\end{center}
\caption{A typical SCADA system}
\label{fig-SCADA}
\end{figure}

SCADA systems (Figure \ref{fig-SCADA}) that tie together decentralized facilities such as power, oil and gas pipelines as well as water distribution and wastewater collection systems, were designed to be open, robust, and easily operated and repaired, but not necessarily secure. Cyber-attacks against these systems are considered extremely dangerous for critical infrastructure (CI) operation and must be addressed in a specific way \cite{zhu2011taxonomy}.

The one-class classification problem is a special case of the conventional two class classification problem, where only data from one specific class are available and well represented. This class is called the target class. Another class, which
is called the outlier class, can be sampled very sparsely, or even not at all. This smaller class contains data that appear when the operation of the system varies from normal, due to a possible attack. OCSVM \cite{jiang2013anomaly} possesses several advantages for processing SCADA environment data and automates SCADA performance monitoring, which can be highlighted as:

\begin{itemize}

\item In the case of SCADA performance monitoring, which patterns in the data are
normal or abnormal may not be obvious to operators. Since OCSVM does not
require any signatures of data to build the detection model it is well suited for
intrusion detection in SCADA environments.

\item Since the detection mechanism does not require any prior information of the
expected attack types, OCSVM is capable of detecting both known and unknown (novel) attacks.

\item In practice, training data taken from SCADA environment could include noise 
Samples and most of the classification based intrusion detection methods are very
sensitive to noise. However, the OCSVM detection approach is robust to noise
samples in the training process.

\item Algorithm configuration can be controlled by the user to regulate the
percentage of anomalies expected.

\item  Due to the low computation time, OCSVM detectors can operate fast enough
for on line SCADA performance monitoring.

\item  Typical monitoring of data of SCADA systems consists of several attributes
and OCSVM is capable of handling multiple attributed data.

\end{itemize}

\section{IT-OCSVM detection mechanism}\label{sec-itocsvm}

The main purpose of the {\it IT-OCSVM} detection mechanism is to perform anomaly detection in a time-efficient way, with good accuracy and low overhead,  within a temporal window adequate for the nature of SCADA systems. In order to achieve the aforementioned goals several operation stages need to be carried out.

\begin{itemize}
\item {\bf Pre-processing} of raw input data in order to feed the {\it IT-OCSVM} module. The attributes in the raw data of the testbed contain all forms: continuous, discrete, and symbolic, with significantly varying resolution and ranges. Most pattern classification methods are not able to process data in such a format. Hence, pre-processing is required before pattern classification models can be built. This consists of two steps: the first involves mapping symbolic-valued attributes to numeric-valued attributes and second is implemented scaling. Different pre-processing techniques are used based on the characteristics of each feature type \cite{lazarevic2003comparative}.

\item {\bf Selection of the most appropriate features} for the training of the {\it IT-OCSVM} model. These features are divided into content and time based features. Since the majority of DoS and probing attacks may use hundreds of packets, time-based features are mostly used. 

\item{\bf Creation of cluster of OCSVM models} that are trained on discrete sources. There are many slow probing attacks that scan the hosts using much larger intervals, thus being able to merge into the overall traffic in the network. As a consequence, these attacks cannot be detected using derived time based features and in order to capture them, the raw data after arriving in the module is split into different datasets according to the sender of the packet. An OCSM module is created and trained for each split dataset. The cluster of split OCSVMs run in parallel with the central OCSVM and produce errors targeted to the specific source.

It is important to mention that a split OCSVM is not created for each source, but only for those sources that produce high traffic in the network during the training period. In order to separate significant nodes a threshold $P_{packets}$ is used and very source that produces a number of packets over this threshold during the training period is marked as a significant node. If during the testing period a source is showing big activity, while not being marked as significant, a medium alarm is fired for it. 

\item {\bf Testing of the traffic dataset} that contains malicious attacks. Based on the models created from the training phase the new dataset is tested against normal patterns. Each OCSVM module returns a function {\it f} that takes the value +1 in a region capturing normal data points (i.e. for events implying normal behaviour of the SCADA system) and takes a negative value elsewhere (i.e. for events implying abnormal behaviour of the SCADA system).

\item {\bf Ensemble of classifiers} The initial outcomes of the different OCSVM modules are combined by the ensemble based mechanism that uses mean majority voting.

\item{\bf Social analysis} Social network analysis is executed based on the network traces and Spearman's rank correlation coefficient is used in order to add weight to alerts produced from different sources.

\item{\bf Fusion of the information}  Due to the possible existence of multiple anomalies in a SCADA system, the outcomes of the different models are gathered and classified in terms of importance. This importance is derived through aggregation and k-means clustering of the outputs.

\item {\bf Communication of the mechanism}. In order to cooperate with other components of the IDS the mechanism sends IDMEF files. The created files describe the nature of the alert, in terms of importance, the position in the system, time etc.

\end{itemize}

\subsection{Ensemble system}\label{sec-ensemble}

\begin{figure}[!hbt]
\begin{center}

\includegraphics[width=0.65\textwidth,natwidth=420,natheight=400]{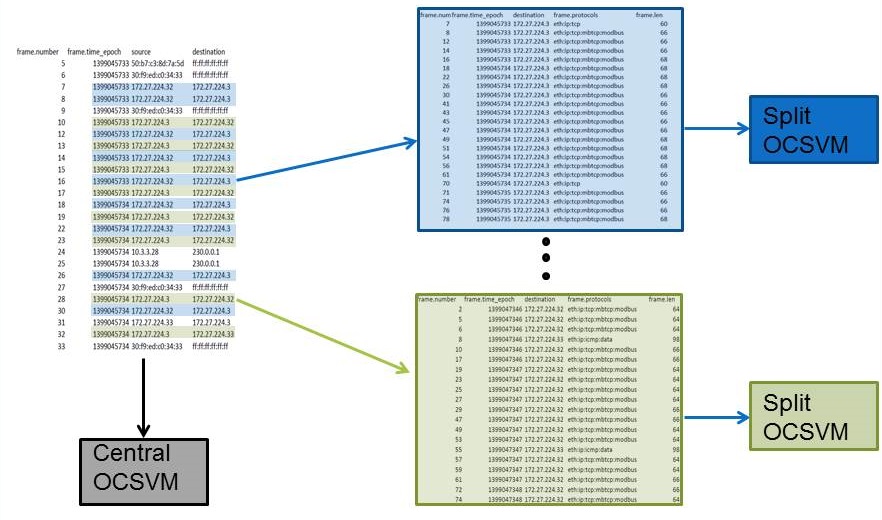}

\end{center}
\caption{Decomposition of the traffic dataset to disjoint subsets}
\label{fig-datasets}
\end{figure}

Ensemble methods can be differentiated according to the extent each classifier affects the others. This property indicates whether the classifiers are dependent or independent. The first situation occurs when the outcomes of a certain classifier affects the creation of the next \cite{menahem2013combining} and the latter, when each classifier is built independently and their results are combined in some fashion \cite{tax2001combining}.
Our proposed mechanism uses an independent ensemble mechanism. Apart from the central OCSVM, which is trained on the entire dataset, a cluster of  split OCSVMs  is automatically created though the decomposition of the traffic dataset into disjoint subsets \ref{fig-datasets}. 
The idea is based on the work of the authors in \cite{chawla2004learning}, who achieved high accuracy by building thousands of classifiers trained from small subsets of data in a distributed environment and through this decomposition high diversity of the methods is achieved. Diversity is an essential feature of an ensemble mechanism in order to achieve high accuracy \cite{tumer1996error}. 

\begin{figure}[!hbt]
\begin{center}

\includegraphics[width=0.65\textwidth,natwidth=420,natheight=400]{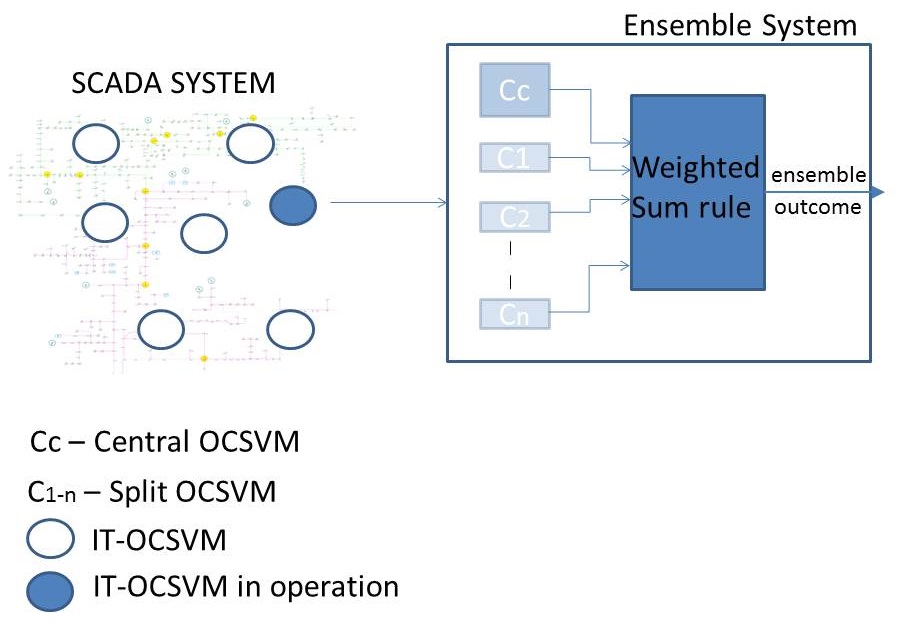}

\end{center}
\caption{Ensemble based system}
\label{fig-ense}
\end{figure}

A combination procedure is then applied so as to produce a single classification for a given instance. There exist many ensemble classification methods, e.g. majority voting, performance weighting, distribution summation an order statistics. These methods have some advantages and disadvantages in terms of accuracy and computational cost. In a real time system, both parameters are crucial and for this reason a simple but effective ensemble mechanism must be chosen \cite{rokach2010pattern}. In order to achieve a balance between these two parameters the outcomes of both the central and the split OCSVMs are combined using a simple algebraic  weighted sum rule using Equation \ref{eq_ens}. The ensemble based mechanism is represented in Figure \ref{fig-ense}. 

\begin{equation}\label{eq_ens}
q_{e}(i,j) = \sum_{n=1}^{N} w_id_{t}(i,j) 
\end{equation}
where, $d_{t}(i,j) $ is the outcome of each individual classifier $n$ for sample data $i$ that originates from node $j$ and $w_i$ is the weight given to each classifier. 

The values that are produced from this stage are weighted using the social analysis module, according to the Spearman's rank coefficient for every significant source. These values are then fed into the aggregation module.

\subsection {Social analysis module}\label{sec-social}

Deviation from the normal protocol operation in communication networks has received considerable attention from the research community in recent years. A malicious node may use the vulnerabilities of the system architecture to perform different kinds of attacks.  Hence, for network reliability it necessary to develop an efficient technique to detect misbehaving clients in a timely manner and the correlation coefficients between entities can effectively detect malicious nodes \cite{hamid2008misbehavior}.

In order to enhance the performance of the OCSVM module, parallel statistical algorithms are executed. For each significant source that is detected during training, a list of the protocols used is created, which are ranked and stored along with the IP/ MAC addresses of each source. During the testing phase the same procedure is executed, thus producing a list of the protocols used for each significant source. The two lists are compared using Spearman's correlation coefficient \cite{spearman1904proof}.

\begin{equation}
\label{eq-coef}
p= 1 - \frac{6\sum d_{i}^{2}}{n(n^2-1)} 
\end{equation}

\begin{figure}[!hbt]
\begin{center}

\includegraphics[width=0.65\textwidth,natwidth=420,natheight=400]{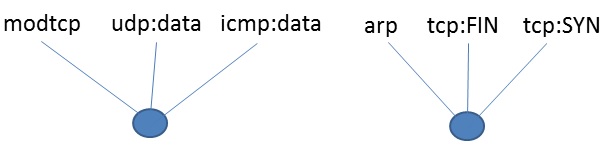}

\end{center}
\caption{Most used protocols used by a node during normal (left) and abnormal (right) operation}
\label{fig-social}
\end{figure}

The final output is a number $q_{j}$ for each source $j$  that indicates whether there is a differentiation in the way that each source behaves during the training and testing period. The output of the method is a value between 1 and 0, with a number close to 1 indicating similar behaviour of the source both in the training and testing sessions. This value is used in order to give additional significance to alerts produced from the specific source from the OCSVM modules (see Equation \ref{eq_wei}. In Figure \ref{fig-social} an example of the protocols that are mainly used during normal and abnormal operation of a node is presented. 

\begin{equation}\label{eq_wei}
 q_{s}(i,j)=\frac{q_{e}(i,j)}{p_j} ,\forall \ q_{e}(i,j) \ with \ source \ node \ {j}
\end{equation}

\subsection {Fusion of alarms - final outputs}\label{sec-fusion}

The OCSVM module produces one initial alarm for any deviation in network traffic from normal, based on the training models that it has created. In real time systems in addition to fast response and accuracy limited communication between the detection modules is desirable. For this reason we implement a fusion procedure which groups alerts per source node and gives final scores to aggregated alerts based on the initial values and the number of similar initial alerts. 

The first stage of fusion consists of an aggregation mechanism that groups individual alerts according to their origin. Each alert $i$  has an initial value $q_i$ (and a source $j$)  based on the procedures described in the previous sections \ref{sec-ensemble},\ref{sec-social}.
Using equations \ref{eq_aggregation} each aggregated alert $j$ is assigned two values $qa_j$ and $qb_j$ that represent its severity. This severity comes from both the sum of the values of the initial alerts and the number of  attacks that originate from the same node.

\begin{equation}\label{eq_aggregation}
q{a_j} =\sum_{i} q_{s}(i,j) , \quad\    q{b_j} =\sum_{i} 1, \forall \ q_{s}(i,j) \ with \ source \ node \ {j}
\end{equation}

During the second stage of fusion the system uses K-means clustering so as to divide the alarms into: possible, medium and severe. The k-means clustering algorithm is one of the simplest and most commonly used clustering algorithms. It is a partitional algorithm that heuristically attempts to minimize the sum of squared errors.

\begin{equation}\label{eq_means}
SSE =\sum_{k=1}^{K}\sum_{j=1}^{N_k}||q{a_j}-\mu_k||^2
\end{equation}
where, $N_k$ is the number of instances belonging to cluster k and $\mu_k$ is the mean of cluster $k$, calculated as the mean of all the instances  belonging to the cluster $i$

\begin{equation}\label{eq_means2}
\mu_{k,i} =\frac{1}{N_k}\sum_{q=1}^{N_k}q{a_q},j \ \forall j
\end{equation}

\begin{figure}[!hbt]
\begin{center}

\includegraphics[width=0.65\textwidth,natwidth=420,natheight=400]{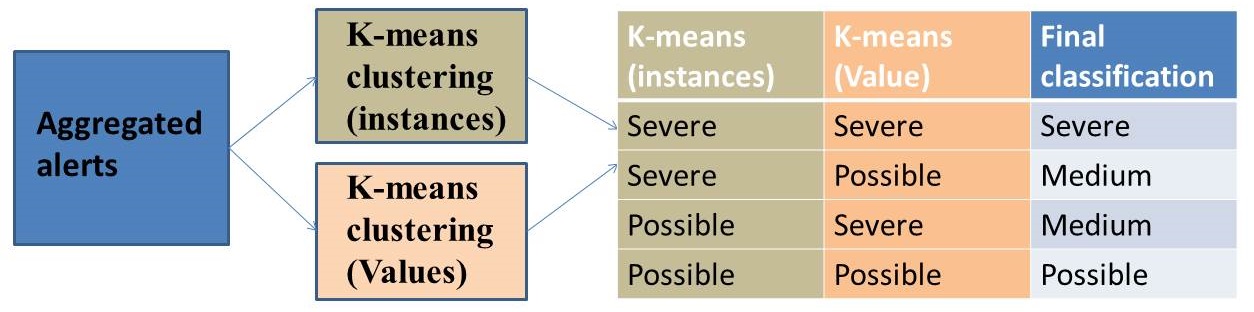}

\end{center}
\caption{K-means clustering module}
\label{fig-kmeans}
\end{figure}

The algorithm begins with an initial set of cluster centers and in each iteration each instance is assigned to its nearest cluster center according to the euclidean distance between the two.
We use two k-means algorithms that run in parallel  which partition alerts into two categories: possible and severe. 
These decisions are then combined according to the above equations and are assigned their final classification: possible, medium or severe, according to Figure \ref{fig-kmeans}. 

\begin{figure}[!hbt]
\begin{center}

\includegraphics[width=0.65\textwidth,natwidth=420,natheight=400]{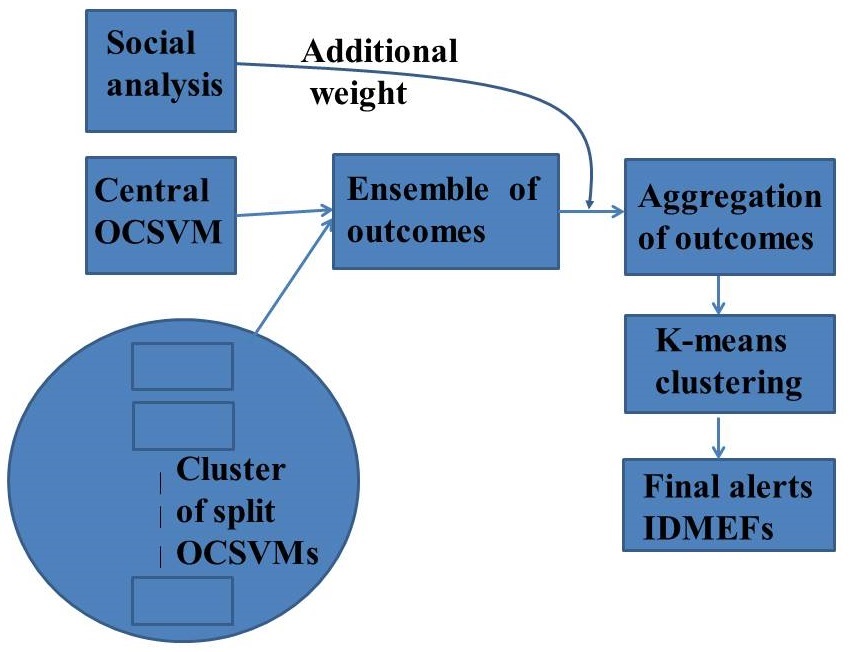}

\end{center}
\caption{Architecture of detection mechanism}
\label{fig-frel}
\end{figure}

Using the fusion procedure, the final alarms that the system produces are significantly decreased and all the sources with suspect behavior are reported. The whole procedure is described in Figure \ref{fig-frel}.

\begin{figure}[!hbt]
\begin{center}

\includegraphics[width=0.65\textwidth,natwidth=420,natheight=400]{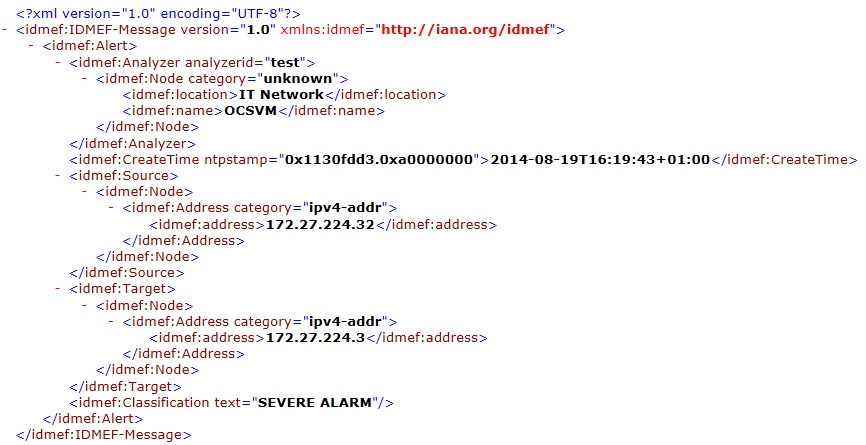}

\end{center}
\caption{Typical IDMEF message produced by the IT-OCSVM mechanism}
\label{fig-idmef}
\end{figure}

In order to cooperate with the other modules the {\it IT-OCSVM} mechanism is integrated in the {\it PID} system and communicates with
the other modules created under the "Cockpit CI project" using IDMEF \cite{debar2007intrusion} files. The IDMEF  defines the experimental standard for exchanging intrusion detection related events and a typical IDMEF file produced by our system is shown in Figure \ref{fig-idmef}. The IDMEF message contains information about the source of the intrusion, the time of the intrusion detection, the module that detected the problem and a classification of the detected attack.  Knowledge of the source node where the intrusion originates is a very important feature an IDS system must have.
Once the infected node is spotted the infection can be limited by the isolation of this node from the rest of the network. Fast and accurate detection of the source node of a contamination is crucial for the correct function of an IDS.

\subsection{Nature of the trial}\label{sec-trials}
\begin{figure}[!hbt]
\begin{center}

\includegraphics[width=0.450\textwidth,natwidth=420,natheight=400]{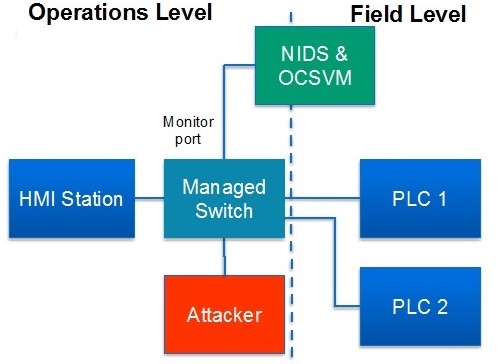}
\includegraphics[width=0.450\textwidth,natwidth=420,natheight=400]{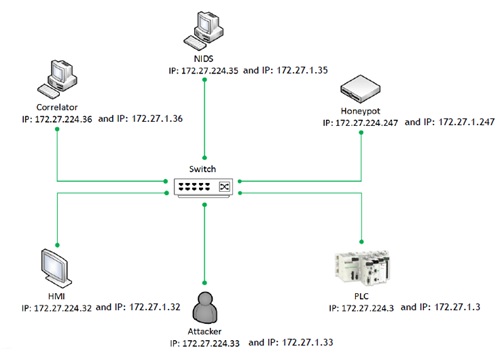}

\end{center}
\caption{Architecture of the testbed}
\label{fig-testbed}
\end{figure}

The trial is conducted off line with the use of two datasets extracted from the testbed (Figure \ref{fig-testbed}. 
The testbed architecture mimics a small-scale SCADA system, comprising the operations and field networks, which include a Human-Machine Interface Station (for process monitoring), a managed switch (with port monitoring capabilities for network traffic capture), and two Programmable Logic Controller Units, for process control.  The {\it NIDS} and {\it IT-OCSVM} modules are co-located on the same host, being able to intercept all the traffic flowing in the network.

Three kinds of attacks are evaluated:

\begin{itemize}
\item	{\bf Network scan attack} 
In a typical network scan attack, the attacker uses TCP/FIN scan to determine if ports are closed to the target machine; closed ports answer with RST packets while open ports discard the FIN message. FIN packets blend with background noise on a link and are hard to detect.
\item	{\bf ARP cache spoofing  - MITM attack}
{\bf ARP cache spoofing} is a technique where an attacker sends fake ARP messages. The aim is to associate the attacker's MAC address with the IP address of another host, thus causing any traffic meant for that IP to be sent to the attacker instead. The attacker could choose to inspect the packets, modify data before forwarding ({\bf  man-in-the-middle attack}) or launch a denial of service attack by causing some of the packets to be dropped. 
\item	{\bf DoS attack}
Network flood refers to the situation where the attacker floods the connection with the PLC by sending SYN packets. 
\end{itemize}

\section{Results and analysis}\label{sec-evaluation}

Each OCSVM that is trained produces a separate model file which is used in order to classify new data as normal or malicious. All OCSVMs use Gaussian RBF Kernel functions with default parameters $\sigma =0.01$  and $ \nu =0.001$. 
The parameters used during evaluation of the proposed IT-OCSVM detection method are shown in table \ref{tab-param}.

\begin{table}[!hbt]
\caption{Evaluation  parameters}
\begin{center}

\begin{tabular}{|c@{ }|c@{ }|}\hline
 {\small \it $Indepedent Paramater$}  & {\small \it Default value}  \\\hline\hline
$\nu$  & $0.001$\\\hline
$\sigma$  & $0.01$ \\\hline
$P_{packets} \  - \ (No \ of \ split \ OCSVMs)$   & $\frac{1}{100}(5)$ \\\hline
$Ensemble \ mechanism$  & $Weighted  \ sum \ value$ \\\hline
$k-means$  & $2 \ stages$ \\\hline
\end{tabular}
\end{center}

\label{tab-param}
\end{table}

\begin{figure}[!hbt]
\begin{center}

\includegraphics[width=0.65\textwidth,natwidth=420,natheight=400]{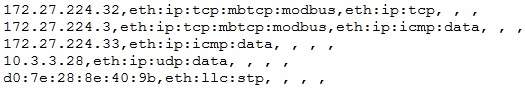}

\end{center}
\caption{Sources file created by the OCSVM module}
\label{fig-sources}
\end{figure}

During execution of the proposed detection mechanism, a separate file that contains information about the split sources is also created. This file is used in order to split the testing dataset according to the sources that are categorized as important during training. A snapshot of the sources file is shown in figure \ref{fig-sources}. For each row, the IP/MAC address of the source is followed by the {\it 5} most used protocols during the training period, which are sorted in descending order. These protocols are compared with those used by the same source in the testing period in order to calculate the Spearman's correlation coefficient value for each separate source.

\subsection{Initial network traffic analysis}\label{sec-initial}

\begin{figure}[!hbt]
\begin{center}

\includegraphics[width=0.65\textwidth,natwidth=420,natheight=400]{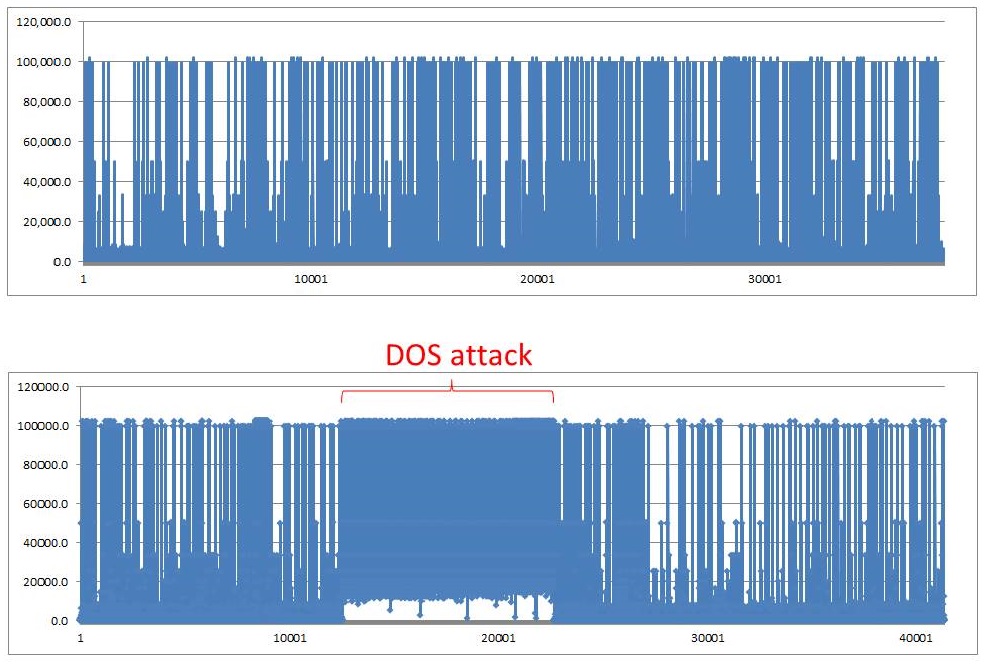}
\end{center}
\caption{Rate of packets during normal and abnormal operation time}
\label{fig-rate}
\end{figure}

In Figure \ref{fig-rate} the rate of the injected packets in the system during normal and abnormal operation of the SCADA system is monitored. In the lower part of the figure, we can observe that when the DOS attack takes place, the rate of the packets injected into the system is much higher compared to a normal operation period.

\begin{figure}[!hbt]
\begin{center}

\includegraphics[width=0.65\textwidth,natwidth=420,natheight=400]{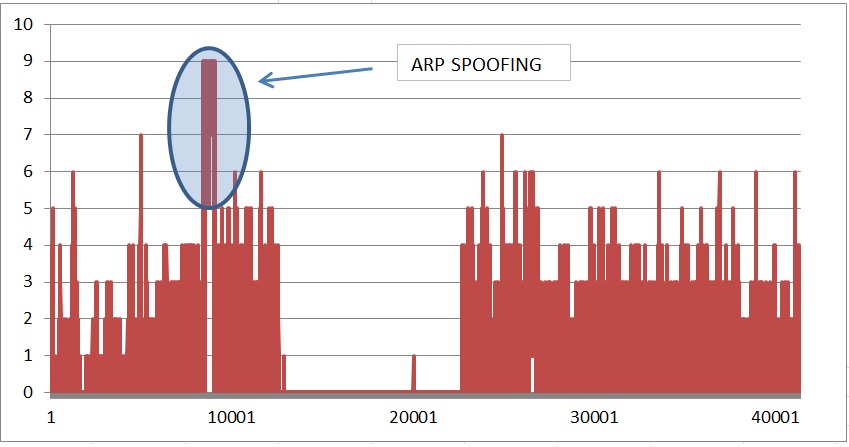}
\end{center}
\caption{ARP spoofing attack - overall dataset}
\label{fig-arp}
\end{figure}

MITM and network scan attacks, on the other hand, do not have such profound consequences on the traffic rate, since they use few messages (MITM) or they merge with the overall traffic. As can be observed from Figure \ref{fig-arp}, an ARP spoofing attack can be identified by using feature {\it 5} in the overall dataset. Apart from the actual attack there are also other instances where the central OCSVM would probably fire an alarm due to high values of this feature.

\begin{figure}[!hbt]
\begin{center}

\includegraphics[width=0.65\textwidth,natwidth=420,natheight=400]{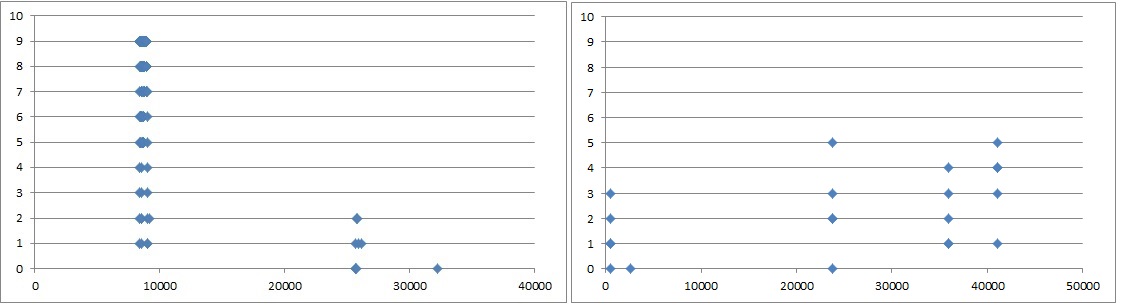}
\end{center}
\caption{ARP spoofing attack - split datasets}
\label{fig-slpit}
\end{figure}

When subsets are used the occurrence of this attack is more obvious. In Figure  \ref{fig-slpit} the values of feature {\it 5} over time for two sources that induce significant traffic in the system are shown. The left diagram presents the number of ARP packets sent by the intruder during the testing period, while the right presents the same distribution for a normal node of the system. The difference in the performance of a node under attack is more evident using the split datasets.

Moreover, the Spearman's rank correlation coefficient of the source that executes a scan attack (Network scan or ARP scan) is heavily affected since node behavior deviates from normal in terms of protocol distribution, thus enhancing the detection capability of our mechanism. 

\subsection{Method evaluation}\label{sec-eval}

This subsection describes the performance of the proposed OCSVM based intrusion detection algorithm for the simulated data.
During the testing period several attack scenarios are simulated in the testbed, which include: a network scan, a network flood and an MITM attack. Since the attacks are performed during different time periods we divide the testing dataset into several smaller ones, each containing a different attack. The testing data consists of normal and attack data and the composition of the data sets is as follows: 

\begin{itemize}
\item Testing set-A' : 1 - 5000: Normal data  
\item Testing set-B' : 5000  - 10000: Normal data + {\bf Arp spoofing} attack + {\bf Network scan}
\item Testing set-C' : 10000 - 25000: Normal data  + {\bf Flooding Dos attack} + {\bf  Network scan} 
\item Testing set-D' : 25000 - 41000:  Normal data  + {\bf MITM attack}
\end{itemize}

The IT-OCSVM method, as shown in \cite{Maglaras_IET_2014}, performs well in terms of detection accuracy and false positive rate \ref{tab-con}. 

\begin{table}[!hbt]
\caption{Performance evaluation of the IT-OCSVM module}
\begin{center}
\begin{tabular}{|c@{ }|c@{ }|c@{ }|}\hline
 {\small \it $Data set$} & {\small \it DA}  & {\small \it FAR}  \\\hline\hline
$A$  & $98.81 \%$ & $1.18 \%$\\\hline
$B$  & $94.6 \%$ & $3.25\%$ \\\hline
$C$  & $95.20 \%$ & $1.51 \%$ \\\hline
$D$  & $96.37 \%$ & $2.3 \%$ \\\hline
$All$  & $96.3 \%$ & $2.5 \%$ \\\hline
\end{tabular}
\end{center}

\label{tab-con}
\end{table}

\subsection {Ensemble mechanism. Parameter $P_{packets}$}\label{sec-ensmble}

\begin{figure}[!hbt]
\begin{center}

\includegraphics[width=0.65\textwidth,natwidth=420,natheight=400]{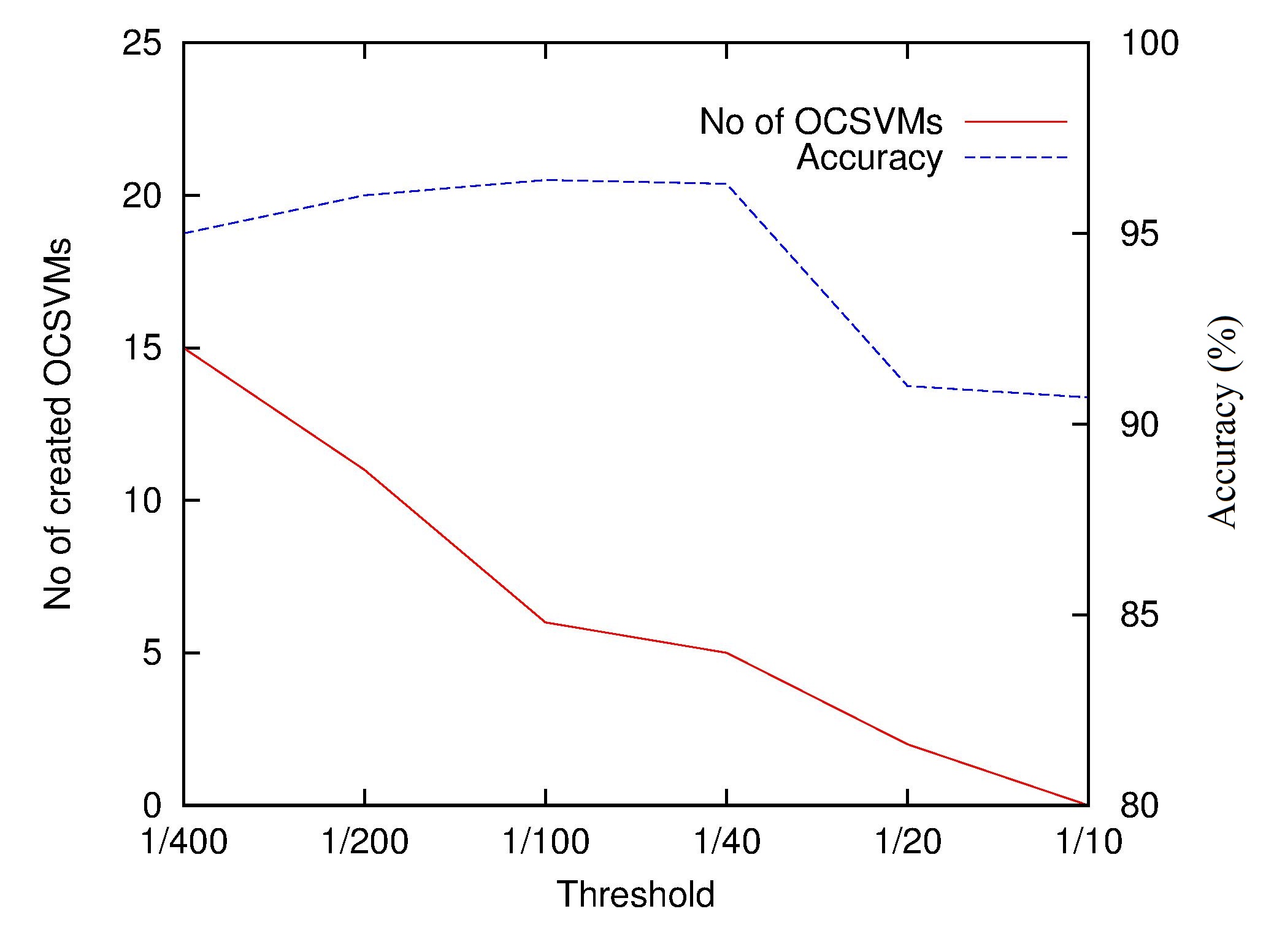}
\end{center}
\caption{Number of automatically created OCSVMs against accuracy of the mechanism}
\label{fig-param}
\end{figure}

The ensemble module affects the performance of the IT-OCSVM. Discrete OCVSMs that are created by the mechanism have an impact on the accuracy of the detection mechanism. 
Regarding which, the cluster of automatically produced OCSVMs can be significantly large(one OCSVM per source) or very small (a total of one or two OCSVMs) according to the value of the parameter $P_{packets}$. Figure \ref{fig-param} shows the 
relation between the value of the threshold parameter  $P_{packets}$ and the number of created OCSVMs as well as how this affects the performance of the detection mechanism.

The parameter $P_{packets}$ is given a value as a proportion of the number of rows of the entire training dataset. According to Figure \ref{fig-param}, it is evident that an appropriate selection of parameter $P_{packets}$ is essential for 
the correct operation of the proposed mechanism. When the parameter is given a very big value then the created OCSVMs are too few to improve accuracy and the performance of the mechanism deteriorates. When this value is 
extremely big  {\it $ \frac{1}{10}$} then the mechanism degrades to a simple OCSVM module. For values relative low the number of created OCSVMs grows without an improvement in systems accuracy. As observed in Figure  \ref{fig-param},  when 
the number of created OCSVMs is relatively large the accuracy of the mechanism drops.
As split training sets are created for sources with no significant traffic in the training period, the OCSVMs that are created are not correctly trained and instead of helping they harass the decision procedure. 
For values of the threshold parameter between {\it $ \frac{1}{40}$} and {\it $\frac{1}{100}$} the accuracy of the detection mechanism is optimal. In fact, for these values, the accuracy of the IT-OCSVM mechanism compared to the initial OCSVM is 
{\it 6 \%} better. That is, the simple OCSVM achieves accuracy of {\it 90.7 \%} while the proposed IT-OCSVM reaches {\it 96.4\%} (Figure \ref{fig-param}). Apart from the profound improvement in terms of accuracy the proposed mechanism has other advantages compared to a simple OCSVM module. It creates a decreased number of alarms and also categorizes these according to their severity, as described in the next subsection.

\subsection {Impact of the fusion mechanism}\label{sec-impact}

The  fusion of the alarms produced by the individual OCSVMs consists of two stages. The first is aggregation of alarms per source and the second is the clustering of them using a two stage k-means clustering algorithm, as described in section \ref{sec-fusion}. 
The outcome of the fusion procedure is the deduction of the communicated alarms in the system and also their classification as: possible, medium and severe. In Table \ref{tab-fus}  the number of initial and final aggregated alarms is presented. It can be observed that the number of final alarms is significantly lower compared to the initial ones, thus reducing the communication costs that such a distributed mechanism have in the network. 

\begin{table}[!hbt]
\caption{Aggregated alarms produced by IT-OCSVM are significantly decreased compared to the initial ones}
\begin{center}
\begin{tabular}{|c@{ }|c@{ }|c@{ }|}\hline
 {\small \it $Data set$} & {\small \it Initial alarms}  & {\small \it Aggregated alarms}  \\\hline\hline
$A$  & $129 $ & $16$\\\hline
$B$  & $658 $ & $21$ \\\hline
$C$  & $9273 $ & $18$ \\\hline
$D$  & $203 $ & $16$ \\\hline
$All$  & $10507 $ & $22$ \\\hline
\end{tabular}
\end{center}

\label{tab-fus}
\end{table}

\begin{figure}[!hbt]
\begin{center}

\includegraphics[width=0.45\textwidth,natwidth=420,natheight=400]{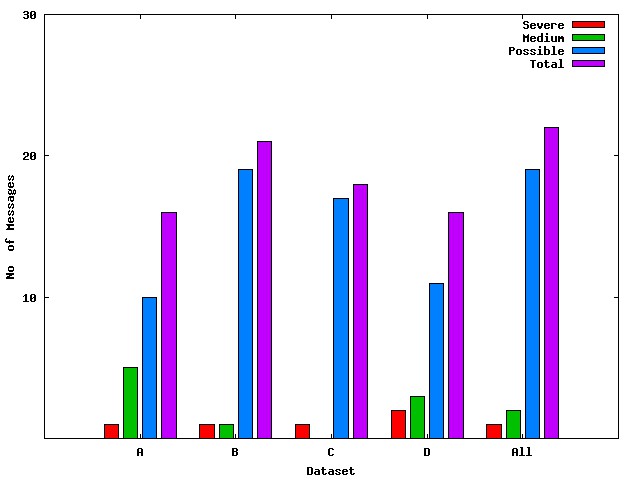}
\includegraphics[width=0.45\textwidth,natwidth=420,natheight=400]{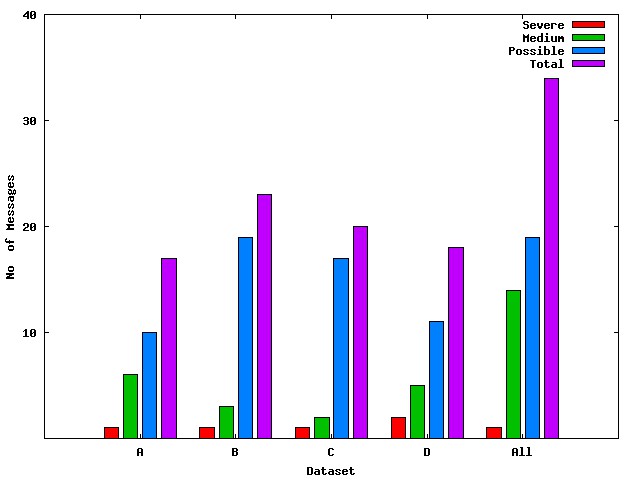}
\end{center}
\caption{IT-OCSVM categorizes aggregated alarms. The left diagram shows aggregated alarms created by IT-OCSVM without the additional medium alarms and the right diagram illustrates all the final alarms created by the IT-OCSVM }
\label{fig-fus}
\end{figure}

The IT-OCSVM system categorizes alarms according to the level of severity they have, with most being classified as possible and those few originating from real attacks in the system are termed severe (see Figure \ref{fig-fus}). 
Since the proposed mechanism is part of a distributed PIDS, the information sent by the IT-OCSVM can be combined with those sent by the other detection modules. 
For this reason this categorization of the alarms is important, if  correct  final decisions about the situation in the network are to be taken.

\subsection{Computational cost and time overhead}\label{sec-over}

The complexity of an IDS can be attributed to hardware, software and operational factors. For simplicity, it is usually estimated as the computing time required to perform classification of the dataset and output the final alarms.
Increasing the number of classifiers usually increases the computational cost and decreases their comprehensibility. For this reason, special care must be taken when choosing parameter $P_{packets}$. 
As mentioned in subsection \ref{sec-ensmble}, this parameter determines the number of created split datasets and thus, split OCSVMs. 
While the increase in number of split OCSVMs does not impose any significant increase in the method's performance, this may slow down the detection mechanism.
 
\begin{figure}[!hbt]
\begin{center}

\includegraphics[width=0.65\textwidth,natwidth=420,natheight=400]{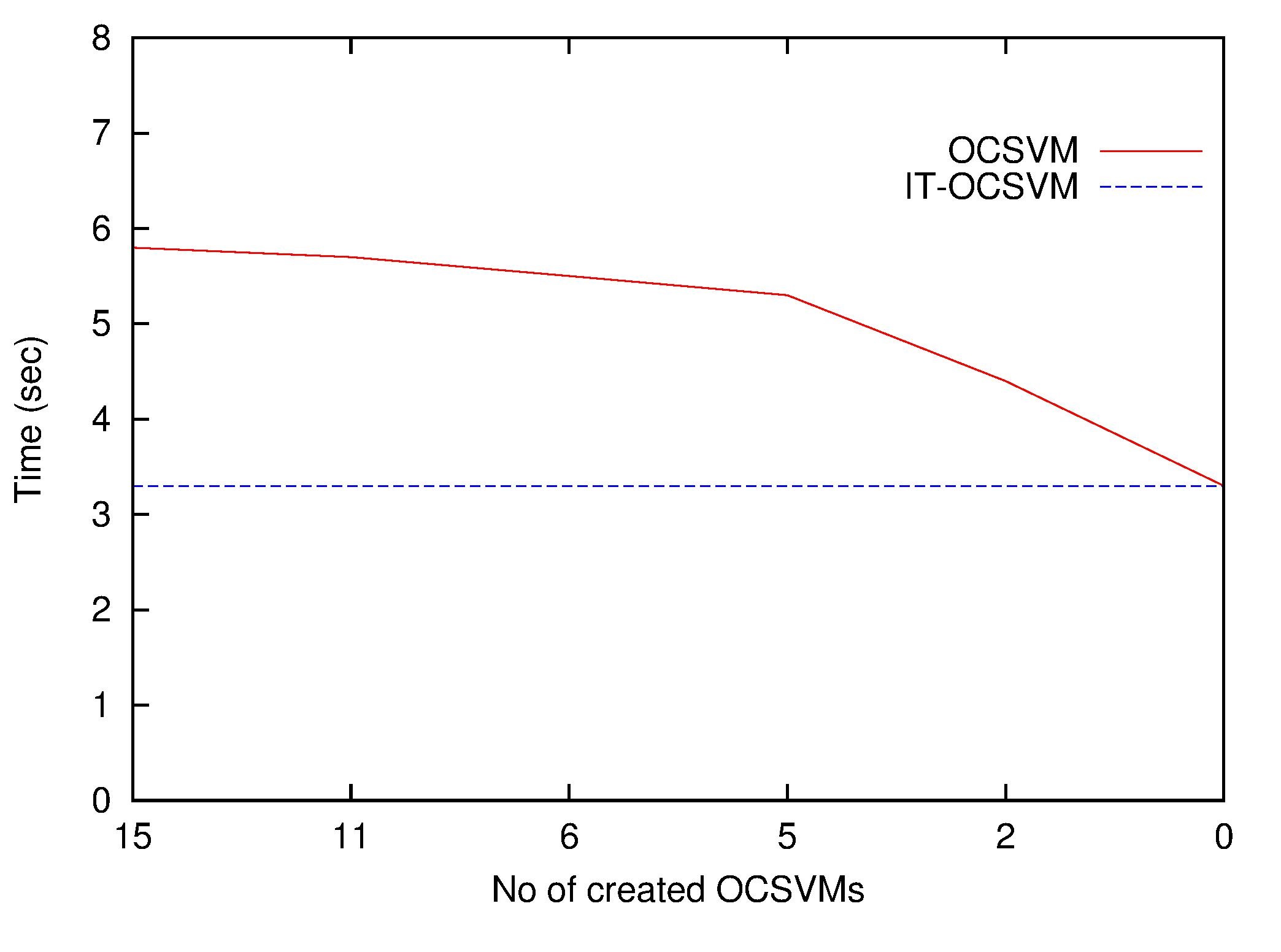}
\end{center}
\caption{Approximate execution time for the entire testing dataset}
\label{fig-tim}
\end{figure}

In figure \ref{fig-tim}, we illustrate the time performance of the method compared to a simple OCSVM. The evaluation was conducted on a PC with Intel 
core 2 duo 1.7 Mhz CPU, 2GB main memory, 80GB hard disk 7200 rpm hard disk and Microsoft windows 7 64bit.

According to Figure \ref{fig-tim}, the execution time of the proposed IT-OCSVM is bigger compared to a simple OCSVM method.
However,or the extreme configuration where {\it 15} OCSVMs are created the performance gap increases and the proposed IT-OCSVM is {\it 75 \%} slower than the simple OCSVM, whereas when the IT-OCSVM operates under the default configuration ({\it 5-6} OCSVMs) the performance gap is  {\it 55 \%}.
Based on these observations we conclude that the system, under the default configuration, performs a classification in a comparable time to that of a simple OCSVM classifier and thus, it can be adopted in soft real-time applications.

We have to mention that the performance evaluation conducted in this subsection, does not include the time that each detection mechanism needs in order to create and disseminate IDMEF messages or the computational and time costs that the correlator needs in order to collect and analyze the alarms. It is evident that the OCSVM classifier, compared to the proposed IT-OCSVM, needs significant additional time in order to send all the detected alarms. Moreover, the categorization that is performed by the IT-OCSVM mechanism reduces the computational load of the correlator that collects the alarms from the distributed detection agents.

\section{Conclusion}\label{sec-concl}
\label{sec-conclusions}

This article analyzes the IT-OCSVM mechanism and evaluates it against the baseline method for different attack scenarios. The detection mechanism, which runs in a distributed way, can be used in large SCADA networks with no additional modifications. The combination of social network analysis metrics with machine learning classification techniques,  enhances the performance of the detection mechanism and improves accuracy for all the simulation scenarios investigated. Moreover, the aggregation procedure embedded in the proposed mechanism decreases the overhead of the IT-OCSVM and makes it easily incorporable in a soft real time system. That is, the method produces a small amount of final alerts and manages to detect all the simulated attacks.

In future work the proposed mechanism will be further enhanced in order to decrease false alarms and increase detection accuracy. It will be tested in a bigger hybrid testbed under different attack scenarios an other behavior patterns will be also evaluated, e.g. the patterns of interaction among entities, thereby adding more sophistication to the detection mechanism.  

\bibliographystyle{elsarticle-harv}

\bibliography{ocsvm}
\end{document}